\documentclass[sigconf]{acmart} 

\AtBeginDocument{
  }

\renewcommand\footnotetextcopyrightpermission[1]{}

\usepackage{enumitem}
\usepackage{adjustbox}
\usepackage{tabularx}
\usepackage{verbatim}
\usepackage{fancyvrb}
\usepackage{fvextra}  

\DefineVerbatimEnvironment{Highlighting}{Verbatim}{breaklines,commandchars=\\\{\}}

\begin{document}

\title{Rewriting Video: Text-Driven Reauthoring of Video Footage}

\author{Sitong Wang}
\affiliation{
  \institution{Columbia University}
  \city{New York}
  \state{NY}
  \country{USA}
}
\email{sw3504@columbia.edu}

\author{Anh Truong}
\affiliation{
  \institution{Adobe Research}
  \city{New York}
  \state{NY}
  \country{USA}
}
\email{truong@adobe.com}

\author{Lydia B. Chilton}
\affiliation{
  \institution{Columbia University}
  \city{New York}
  \state{NY}
  \country{USA}
}
\email{chilton@cs.columbia.edu}

\author{Dingzeyu Li}
\affiliation{
  \institution{Adobe Research}
  \city{Seattle}
  \state{WA}
  \country{USA}
}
\email{dinli@adobe.com}

\renewcommand{\shortauthors}{Wang, et al.}

\begin{abstract}
Video is a powerful medium for communication and storytelling, yet reauthoring existing footage remains challenging. 
Even simple edits often demand expertise, time, and careful planning, constraining how creators envision and shape their narratives. 
Recent advances in generative AI suggest a new paradigm: what if editing a video were as straightforward as rewriting text? 
To investigate this, we present a tech probe and a study on text-driven video reauthoring. 
Our approach involves two technical contributions: (1) a generative reconstruction algorithm that reverse-engineers video into an editable text prompt, and (2) an interactive probe, Rewrite Kit, that allows creators to manipulate these prompts. 
A technical evaluation of the algorithm reveals a critical human-AI perceptual gap. 
A probe study with 12 creators surfaced novel use cases such as virtual reshooting, synthetic continuity, and aesthetic restyling. 
It also highlighted key tensions around coherence, control, and creative alignment in this new paradigm. 
Our work contributes empirical insights into the opportunities and challenges of text-driven video reauthoring, offering design implications for future video tools.

\end{abstract}

\begin{CCSXML}
<ccs2012>
<concept>
<concept_id>10003120.10003121</concept_id>
<concept_desc>Human-centered computing~Human computer interaction (HCI)</concept_desc>
<concept_significance>500</concept_significance>
</concept>
</ccs2012>
\end{CCSXML}

\ccsdesc[500]{Human-centered computing~Human computer interaction (HCI)}

\keywords{Video reauthoring, Text-driven video editing, Generative video models, Creative AI tools}

\begin{teaserfigure}
\centering
  \includegraphics[width=0.95\textwidth]{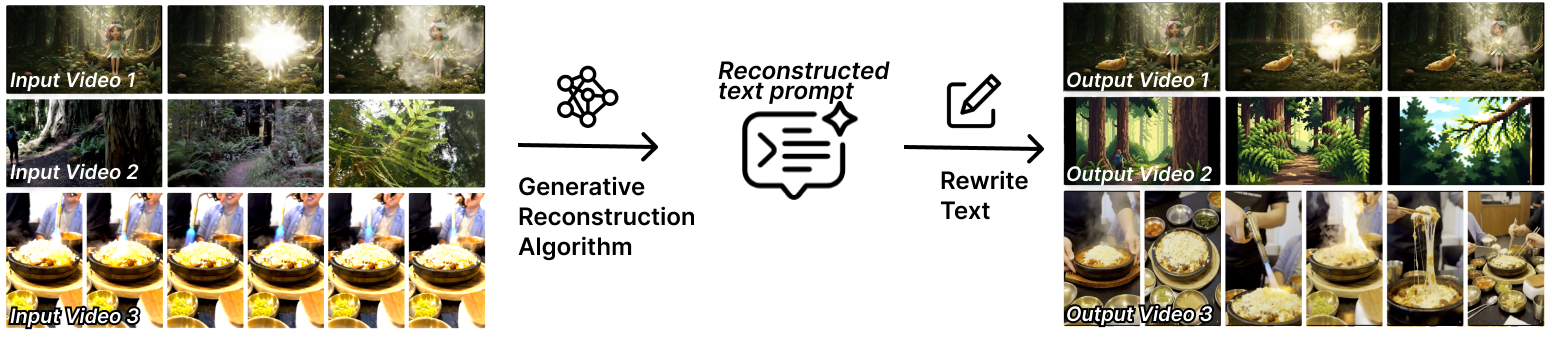}
  \label{fig:teaser}
  \caption{Workflow of our text-as-interface approach for video reauthoring. A generative reconstruction algorithm extracts an editable text representation from an input clip, which can then be modified via text rewriting to generate a new video output. See the complete list of use cases in Table~\ref{tab:use_cases}. Here we highlighted (1) adding a new character; (2) changing the style to pixel art; and (3) diversifying the camera viewpoint.}
\end{teaserfigure}

\maketitle
\pagestyle{plain} 

\section{Introduction}
Video is one of the most powerful media for communication and storytelling, yet reauthoring or adapting existing video content remains remarkably challenging. 
For professional and personal use alike, seemingly simple editing tasks demand significant technical expertise and time. 
Achieving a smooth transition may require deliberate camera movements during filming; capturing a scene from multiple angles often depends on a complex multi-camera setup. 
A single missed shot can be catastrophic to the intended narrative, and advanced techniques like reference-based editing are inaccessible to everyday creators.
These barriers fundamentally restrict how people can shape video into the narratives they envision.

Recent advances in AI---particularly in video understanding and text-to-video generation---offer a new paradigm for lowering these barriers. 
What if reauthoring a video were as straightforward as editing text? 
This vision recasts video not as a fixed artifact, but as a medium with an exposed, editable ``script.'' 
This script is not merely a transcript of spoken words; it encompasses the intricate layers that constitute a video: its visual compositions, rhythmic pacing, and overall aesthetic. 
This concept moves beyond conventional timeline manipulation or simple transcript-based editing toward a more holistic, semantic form of video authoring.

Unlike simply editing a transcript, which primarily focuses on the spoken word and its textual representation, we treat all modalities of the video as inherently editable components. 
This paradigm empowers creators to manipulate the video's essence with the fluidity of revising a written document.
They could directly alter the visual flow, modify the temporal rhythm, or fine-tune other elements that shape the video's impact. 
The aspiration is to democratize the editing process by making all aspects of video manipulable through a script-like interface, unlocking new possibilities for creative expression and content adaptation.

To explore the potential and challenges of this text-driven approach, we developed a tech probe and conducted a probe study. 
Our work proceeds in three stages:
\begin{itemize} \itemsep0em
\item We first developed a generative video reconstruction algorithm that iteratively reverse-engineers a source video into an editable text prompt. Our evaluations show the algorithm's accuracy converges after 3-6 iterations for a variety of video styles.
\item Our analysis revealed a critical human-AI perception gap: while AI metrics prioritized frame-level fidelity (e.g., object accuracy), human viewers valued temporal-narrative coherence---qualities like pacing, vibe, and smooth transitions.
\item Building on these insights, we designed and built an interactive probe system that exposes the editable prompts and integrates conversational assistance. 
The interface allows users to edit prompts directly and provide visual anchors (e.g., the first frame) to guide the generative process.
\end{itemize}

This work investigates the following research questions:
\begin{enumerate}[label={\textit{RQ{\arabic*}}:}]
\item How do creators envision using text-driven video reauthoring tools, and what novel use cases are unlocked?
\item What challenges and opportunities arise when video editing is mediated through text, and what are the design implications for future systems?
\end{enumerate}

Through a probe study with 12 creators, we identified emerging practices such as video inpainting, B-roll variation, multi-angle synthesis from a single capture, and reference-guided adaptation of style and rhythm.
These findings show how text can operate not just as an interface, but as a creative instrument---enabling creators to reshoot, remix, and restyle footage through language.
Our work also reveals a central tension: while text-driven workflows can democratize video production by lowering technical barriers, they introduce new challenges in ensuring coherence, control, and alignment between human intent and generative output.

Together, these insights highlight the potential and current limitations of text-driven video reauthoring, pointing toward future systems that support more accessible, expressive, and semantically aware forms of video creation. 
The contributions of this paper are:
\begin{itemize}\itemsep0em
\item A generative reconstruction algorithm that enables videos to be deconstructed and reauthored through editable text prompts.
\item The identification and analysis of a human-AI perception gap in video generation, where AI fidelity metrics diverge from human priorities of narrative and temporal coherence.
\item An interactive probe, Rewrite Kit, demonstrating how text-based interfaces can support exploratory and iterative forms of video creation.
\item Empirical insights from a probe study identifying novel practices, challenges, and design opportunities for text-driven video reauthoring.
\end{itemize}

\section{Related Work}
Our work is situated at the intersection of video manipulation, multimodal AI, and text-driven creative interfaces. 
We review these three areas to position our contribution.

\subsection{Creative Manipulation of Videos}
Video manipulation has long been central to creative production. 
Previous HCI systems for text-based editing~\cite{rubin2013content, fried2019text, huber2019b, roughcut, podreels} treated transcripts as semantic interfaces for navigating and rearranging footage. 
This approach made video editing more legible by turning it into something ``writable;'' yet its expressivity was largely limited to literal, selective operations on \emph{speech-heavy content}. 
While such tools simplified tasks like selecting engaging clips~\cite{podreels} or inserting B-rolls~\cite{huber2019b}, they offered little capacity to alter camera perspective, scene composition, or visual style---elements core to narrative transformation.

More recent AI systems have expanded this scope. 
Keyframe-guided inpainting methods~\cite{guo2025keyframe} can synthesize motion and composition while preserving temporal coherence, and tools like Runway~\cite{runway_creatingwithaleph} allow users to restyle or extend footage through text prompts. 
However, these systems still tend to operate on a local level, treating video as a sequence of frames to be modified. 
This makes them powerful for substitution or restyling but less suited for enacting higher-level narrative or structural changes. 

Our work builds on this trajectory but shifts the focus from localized editing to holistic, generative rewriting, where text is used to reimagine the video's narrative, pacing, and aesthetic structure.
We explore how creators use language to reshoot, remix, and re-style existing footage.

\subsection{Multimodal Generative Models}
The technical foundation for our work lies in recent advances in multimodal generative models. 
In the image domain, models like CLIP~\cite{radford2021learning} and principles like cycle consistency from CycleGAN~\cite{zhu2017unpaired} established robust connections between vision and language.
These connections enabled unified workflows that integrate multimodal understanding and generation within a shared framework~\cite{xie2025reconstruction}.

In the video domain, this progress is also accelerating. 
Large-scale models like Veo 3~\cite{wiedemer2025video} are demonstrating generalist temporal reasoning, while text-to-video generative systems such as Runway Gen-4~\cite{runway2025gen4} and OpenAI Sora 2~\cite{sora2025openai} achieve impressive generation quality. 
Despite this progress, challenges in long-range continuity and narrative control persist. 

Together, these developments reflect rapid progress in unified multimodal generative models.
Crucially, however, rather than interactive creative use, so far the research focus remains largely model-centric, prioritizing accuracy and synthesis fidelity. 
Our work builds on these advances but shifts the focus toward human-AI co-creation, exploring how generative models can support text-driven video reauthoring.

\begin{figure*}
\centering
\includegraphics[width=0.48\textwidth]{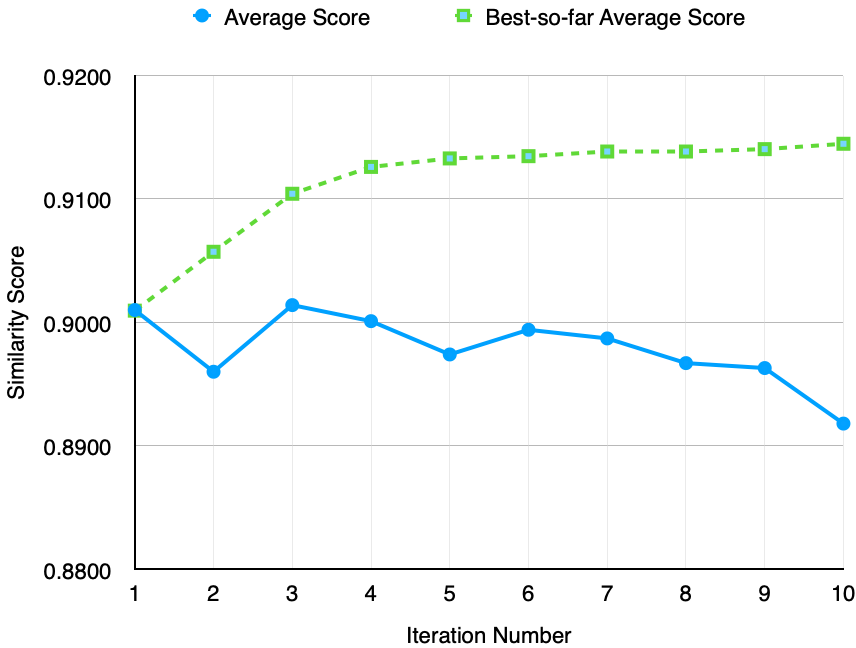}
\caption{
The average similarity across iterations and the rationale for early stopping. 
The solid line shows the mean similarity score if we keep iterating uniformly, while the dashed line shows the ``best-so-far'' average, calculated as if each clip stops the first time it reaches its peak score. Performance gains typically saturate by 3-6 iterations. Further iterations often lower the per-iteration average due to ``prompt drift,'' whereas the best-so-far curve remains flat, demonstrating the benefit of implementing per-clip early stopping.}
\label{fig:iterations}
\end{figure*}

\subsection{Text-Driven Creative Interfaces}
Our interface design builds on HCI research exploring natural language as a co-creative medium. 
In writing, systems like CoAuthor~\cite{lee2022coauthor} have studied the collaborative dynamics between humans and language models. 
In visual art, tools like Promptify~\cite{brade2023promptify} and PromptCharm~\cite{wang2024promptcharm} support the iterative refinement of text-to-image prompts. 
This body of work shows that language can lower creative barriers but also shifts authorship toward a practice of prompting, steering, and interpretation.

Building on this foundation, HCI research has begun articulating design principles for prompt-based creation.
Work such as design guidelines for prompting text-to-image models~\cite{liu2022design} and How to Prompt~\cite{dang2022prompt} frames prompting as a form of creative reasoning---one that involves articulating intent, interpreting model feedback, and negotiating meaning through language.
These insights foreground the expressive and cognitive challenges of text-mediated creation, but prior work has largely centered on static text or image artifacts.

Our work extends this line of inquiry to video, where language functions as an editable substrate for reauthoring and transformation.
Rather than optimizing prompts for fidelity, we investigate how creators rewrite video---using language to author temporal, stylistic, and narrative change.
This perspective situates text-driven video reauthoring within broader efforts to design human-AI co-creative systems that center text as a medium of authorship.

\section{Generative Reconstruction Algorithm}
To enable text-driven video reauthoring, we first proposed a method to deconstruct a source video into a faithful textual representation---an editable prompt that can accurately reproduce the original footage. 
We developed a generative reconstruction algorithm that achieves this through a closed-loop process of prompt generation, video synthesis, and comparative analysis.

\subsection{Iterative Reconstruction Pipeline}
Our pipeline reverse-engineers a video into a prompt through iterative refinement. 
Each cycle consists of generating a video from a candidate prompt, comparing it to the original, and using a vision-language model (VLM) to suggest improvements to the prompt for the next cycle.

\subsubsection{Input and Initialization}
The pipeline takes a video clip as input and uses a VLM (Gemini 2.5 Pro) to generate an initial descriptive prompt. 
The VLM is instructed to create a detailed, temporally-aware prompt suitable for a text-to-video model~\cite{google2025gemini}, based on an analysis of the video at 16 FPS. 
The instruction template is: “Reverse engineer the \{time-duration\} second video to create a clear and descriptive prompt that can be used to reproduce the video with a text-to-video model. Return the prompt only. Include temporal sequencing.”
This output serves as the seed prompt for the first iteration.

\subsubsection{Generation and Comparison}
The candidate prompt is passed to a text-to-video model (Veo 3), conditioned on the first frame of the source video to maintain compositional consistency. 
The resulting synthetic video is then compared against the original. 
The same VLM (Gemini 2.5 Pro) analyzes both videos and generates a structured difference report, highlighting semantic gaps and providing a revised, improved prompt.

\subsubsection{Iterative Refinement}
The revised prompt from the comparison step becomes the input for the next cycle. 
This generation, comparison, refinement loop continues until a convergence criterion is met, such as when similarity scores plateau. 
In our experiments, performance typically peaked within 3–6 iterations (Figure~\ref{fig:iterations}). 
The final output is the prompt that yielded the highest similarity score during this process.

\subsection{Automated Evaluation}
We conducted an automated evaluation to measure the algorithm's convergence behavior and effectiveness.

\subsubsection{Dataset and Metrics}
We curated a dataset of 30 diverse 8-second video clips (the clip length supported by the video generation model in our pipeline at the time of study), spanning genres such as vlogs, animations, cinematic scenes, and social media content. 
To quantify the alignment between a generated video and the original, we computed the frame-aligned cosine similarity between their CLIP (ViT-B/32) embeddings. 
This score serves as our primary metric for reconstruction accuracy.

\subsubsection{Results}
We ran the pipeline for 10 iterations on all 30 clips.
The iterative process successfully improved upon the initial prompt in 80\% of cases (24 of 30). 
The mean similarity score peaked at 0.9145, an average improvement of +0.0135 over the initial prompt's score of 0.9010.

As shown in Figure~\ref{fig:iterations}, performance improved rapidly in early iterations, with most clips reaching their peak similarity between iterations 3 and 6 (mean best iteration = 4.23). 
Beyond this point, gains diminished. Extended refinement often led to ``prompt drift,'' where the model over-optimized for minor details at the expense of the core content. 
By iteration 10, the mean similarity had fallen to 0.8918, below the initial baseline. 
This suggests that early-stopping criteria are crucial for optimal performance. 
Clips with fast or complex motion (e.g., dance videos) proved more challenging, achieving lower peak scores and converging later.

\subsection{Human Evaluation and the Perceptual Gap}
To complement our automated metrics, we conducted a small-scale human evaluation to assess the perceptual quality of the reconstructions.

\subsubsection{Method}
Two independent raters (Cohen’s Kappa $\kappa=0.82$) viewed all 30 pairs of original and best-reconstructed videos. 
They rated how well the reconstruction reproduced the original on a 1-7 scale and described the main differences. 
Raters were compensated \$25 per hour.

\subsubsection{Results: A Human-AI Perceptual Gap}
The mean rating across all clips was 5.07 out of 7, with 96.7\% of reconstructions judged as acceptable ($\ge$4/7). 
However, the qualitative feedback revealed a critical \emph{perceptual gap} between human and AI evaluation.

Our automated CLIP-based metric, like the VLM's comparison reports, emphasized frame-level visual fidelity: object accuracy, color, lighting, and composition. 
In contrast, human evaluators prioritized temporal and narrative qualities: the ``vibe,'' pacing, rhythm, and realism of motion (e.g., choreography, character gait).
Participants would rate a reconstruction lower due to a subtle difference in timing or movement, even when individual frames were visually accurate.

This finding suggests that while current AI models excel at seeing \textit{what} is in a video, human perception hinges on \textit{how} it unfolds over time. 
The essence of a video, from a human perspective, lies in its temporal flow and coherence. 
This highlights a key challenge for future work: developing models and interfaces that can reason about and manipulate the rhythmic and narrative qualities that make video a compelling medium.

\begin{figure*}
\centering
\includegraphics[width=0.9\textwidth]{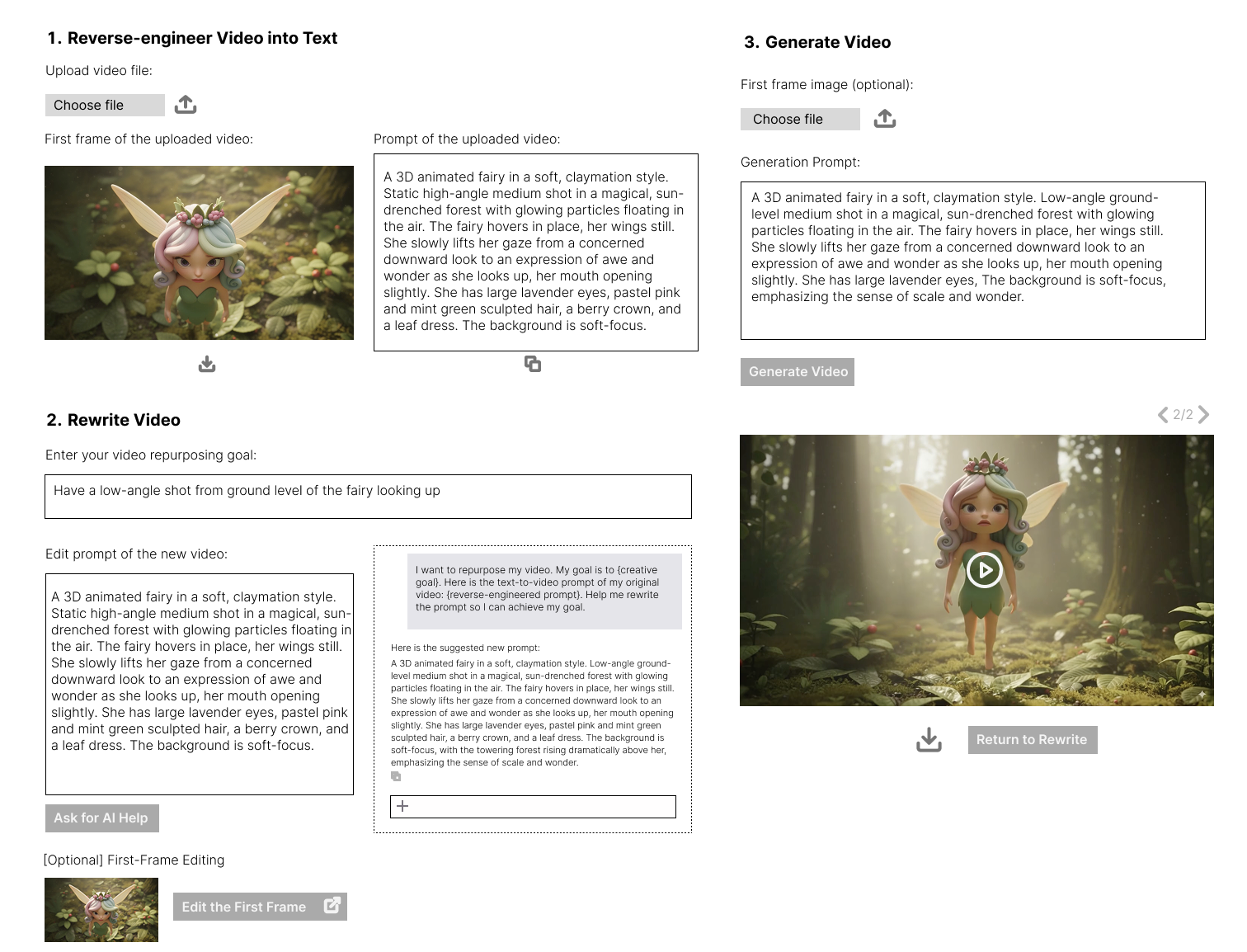}
\caption{The user interface of the Rewrite Kit technology probe. A creator's three-part workflow: reverse-engineer, rewrite, and generate. 
}
\label{fig:probe}
\end{figure*}

\section{Design of the Probe: Rewrite Kit}

To investigate how creators engage with text as an interface for video reauthoring, we developed Rewrite Kit, a web-based technology probe (Figure \ref{fig:probe}). 
By technology probe, we mean a functional but intentionally unfinished system used not to evaluate the tool itself, but to elicit users’ strategies, interpretations, and interactional tensions in practice. 
Accordingly, our design was guided not by the goal of creating a finished or optimal editing tool, but by using the system as a research instrument to surface how creators use text to reimagine video~\cite{hutchinson2003technology, boehner2007hci}.

\subsection{Design Goals}
Our design was shaped by an iterative process involving a research team with expertise in HCI, AI, and filmmaking. We established two primary design goals:

\begin{enumerate}[nosep,leftmargin=*,label={\textit{DG{\arabic*}}}]
    \item \textit{Expose the ``Script'' of a Video}: The system needed to translate the implicit visual and temporal structure of a video into an explicit, editable textual artifact. This would provide a concrete object for creators to manipulate and reason about.
    \item \textit{Support Open-Ended, Exploratory Rewriting}: The interface should impose minimal workflow constraints, allowing creators to freely experiment with different ways of reauthoring their footage. It needed to support both direct manipulation of the text and provide scaffolding for those who might need help articulating their goals.
\end{enumerate}

\subsection{The Rewrite Kit Interface and Workflow}
The probe is organized around a simple three-part workflow: reverse-engineer, rewrite, and generate.

\subsubsection{Reverse-engineering Video into a Textual ``Script''}
Upon uploading a video, Rewrite Kit automatically runs our generative reconstruction algorithm. 
This pipeline reverse-engineers the video into an optimized text prompt that, along with the video's first frame, tries to faithfully reconstruct the original content.
We run the algorithm for six iterations, at which point our technical evaluation showed converging accuracy. 
The resulting prompt is presented to the user in an editable text box next to a thumbnail of the video's first frame (Figure \ref{fig:probe}, Left), fulfilling DG1. 
This prompt becomes the primary medium for creative manipulation.

\subsubsection{Rewriting the Script: Direct and AI-Assisted Prompting}
To support a flexible rewriting process (DG2), the interface provides two modes of interaction. 
Users can directly edit the text prompt in the editor, treating it like any other piece of text.
Alternatively, they can invoke a conversational AI assistant (powered by GPT-5) by clicking ``Ask for AI Help.'' 
This opens a chat window where users can describe their creative goals in natural language. 
To scaffold this interaction, the probe provides a starter message: ``I want to repurpose my video. My goal is to \{creative goal\}. Here is the text-to-video prompt of my original video: \{prompt\}. Help me rewrite the prompt…'' 
Users can then easily copy the AI's suggestions back into the editor or continue the conversation to refine them.

For transformations involving significant stylistic or compositional changes, a revised text prompt alone may not be sufficient. 
Therefore, Rewrite Kit allows users to optionally provide a new first frame as a stronger visual anchor. 
Users can select ``Edit First Frame,'' which uses an image generation model (Gemini 2.5 Flash Image) to create a new starting image based on their textual goals. 
The probe suggests a template for this request: ``I want to repurpose my video. My goal is \{creative goal\}... Suggest an image-editing prompt to get the first frame of my new video.'' 
The user can then use the generated image in the final step.

\subsubsection{Generating and Iterating on the New Video}
Once the user has finalized their rewritten prompt and optional new first frame, they proceed to generation.
This stage uses the Veo3 model to produce the reauthored video. 
The result is displayed in an embedded viewer, with a toggle that allows users to compare different generated versions and their corresponding prompts. 
A ``Return to Rewrite'' button enables a tight iterative loop, allowing users to go back, tweak the text or image, and regenerate as many times as they wish. 

\section{Probe Study Methods}
We conducted a qualitative probe study to understand how creators would engage with a text-driven video reauthoring workflow in practice.

\subsection{Participants}
We recruited 12 video creators (7 female, 5 male; age M=27.4, SD=3.1), each of whom creates or edits videos at least weekly. 
Our recruitment targeted a mix of user groups to gather diverse perspectives:
\begin{itemize}
    \item \textbf{10 novice creators}, representing the primary target users of our tool, were recruited via university mailing lists and word of mouth. All novices had prior experience shooting and editing videos and were proficient with conversational AI tools. Half had previously experimented with text-to-video generation models. Novices were compensated \$25 per hour.
    \item \textbf{2 expert creators} were recruited for their deep domain knowledge. Each had over 10 years of professional video editing experience and extensive familiarity with AI-based generation tools. One expert also had formal training in film production. Experts were compensated \$50 per hour.

\end{itemize}

\begin{table*}[t]
\centering
\begin{adjustbox}{width=1\textwidth}
\renewcommand{\arraystretch}{1.05}
\begin{tabular}{ l  l  l  l  l  l   }
\toprule
\ & User & User goal & Inputs & Desired transformation & Category \\
\midrule
1  & P1  & Change the shooting angle of a dog running & Clip + ref. image & New camera angle & Re-shooting / perspective change \\
2  & P2  & Combine the concert stage and selfie & Two clips & Composited video & Remix / compositing \\
3  & P3  & Add dynamic camera motion to dance footage & Clip & Simulated moving camera & Re-shooting / motion synthesis \\
4  & P4  & Recreate the viral ``historical selfie'' clip for another figure & Ref. clip & Stylized reenactment & Re-worlding / generative narrative \\
5  & P4  & Remove a couple blocking the view in boat footage & Clip & Clean continuous shot & Remix / object removal \\
6  & P5  & Make a YouTube intro like a favorite channel & Ref. clip & Personalized intro & Re-styling / branding \\
7  & P6  & Recreate glitchy hoodie-swap transition & Ref. clip & Viral transition & Re-styling / transition \\
8  & P7  & Add a drone view of the national park & Clip & Synthetic aerial shot & Re-shooting / synthetic viewpoint \\
9  & P8  & Stylize vlog as pixel art & Clip + ref. image & Stylized render & Re-styling / aesthetic transfer \\
10 & P9  & Expand short food clip into rich vlog & Clip + ref. clip & Multi-angle, music-synced vlog & Re-styling / narrative enhancement \\
11 & P10* & Add a yellow slug to the fairy animation & Clip + ref. image & New character inserted & Re-worlding / object addition \\
12 & P10* & Change fairy shot to low angle from ground & Clip & Reframed spatial composition & Re-shooting / scene reframing \\
13 & P10* & Change the camera angle of cat-riding-slug animation & Clip & Alternate camera view & Re-shooting / perspective variation \\
14 & P11* & Generate a smooth transition between two clips & Two clips & Seamless temporal bridge & Remix / transition synthesis \\
15 & P12 & Extend the basketball shot to walking toward the hoop & Clip & AI-generated continuation & Remix / scene extension \\
\bottomrule
\end{tabular}
\end{adjustbox}
\caption{In the probe study, we found 15 potential use cases from 12 participants, spanning a wide diversity of video editing goals, inputs, and desired transformations.}
\label{tab:use_cases}
\end{table*}

\subsection{Procedure}
Each participant attended a one-on-one session lasting approximately 45 minutes. 
Sessions were conducted remotely, and all interactions and dialogues were recorded with participant consent. 
The session was structured in three parts.

First, we introduced the concept of text-driven video reauthoring and provided a brief demonstration of the Rewrite Kit probe's functionalities. 
We explicitly framed the study's goal not as an evaluation of usability or task performance, but as a way to elicit participants' \textbf{explorations, reflections, and aspirations} when reimagining video through text.

Next, we invited participants to engage in an open-ended creative task. 
They were asked to bring their own video clips and brainstorm one or more real-world reauthoring goals they wished to achieve. 
Participants were invited to pursue as many of these goals as time allowed during the session. 
They then used Rewrite Kit freely to work toward producing new videos they would consider ``publishable.''
Participants were encouraged to think aloud throughout this process. 
They were also permitted to use their preferred video editing software for post-processing if they felt it was necessary to finalize their vision.

Finally, each session concluded with a semi-structured interview. 
We asked participants to reflect on their experience, focusing on the challenges they faced, surprising outcomes, and the opportunities they envisioned for such a tool in their creative practice.

\section{Probe Study Findings}

We analyzed the study data using thematic analysis~\cite{braun2006using}. 
The first author conducted an initial coding of the transcripts and session recordings, from which the research team collaboratively developed the final themes.
The findings are organized around our two research questions.

\subsection{RQ1: How do creators envision using text-driven video reauthoring tools? What novel use cases are unlocked?}

Participants used Rewrite Kit to reauthor their videos across a wide range of creative goals, treating text as a medium for \textbf{reshooting}, \textbf{remixing}, \textbf{restyling}, and \textbf{re-worlding} their footage. 
Across the fifteen distinct use cases generated by participants (Table~\ref{tab:use_cases}), creators envisioned and enacted workflows that would traditionally require complex production setups or professional editing tools. 
These practices reveal how text-driven interfaces can unlock new forms of virtual filmmaking, synthetic continuity, aesthetic adaptation, and narrative world-building.

\subsubsection{Text as a Virtual Camera: Re-shooting without Retakes}

\begin{figure*}
\centering
\includegraphics[width=0.73\textwidth]{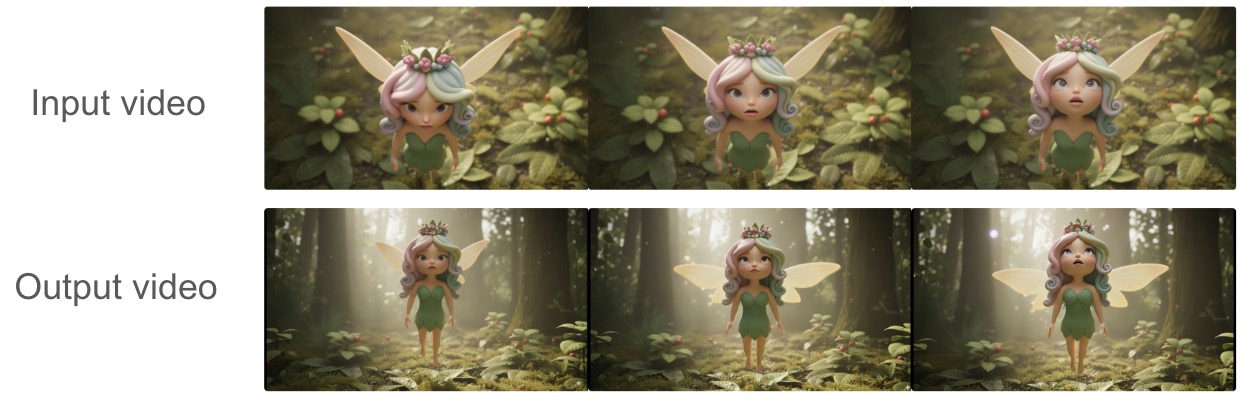}
\caption{Use case 12: Camera angle change. Change the fairy shot to a low angle from the ground.}
\label{fig:use_case_12}
\end{figure*}

\begin{figure*}
\centering
\includegraphics[width=0.73\textwidth]{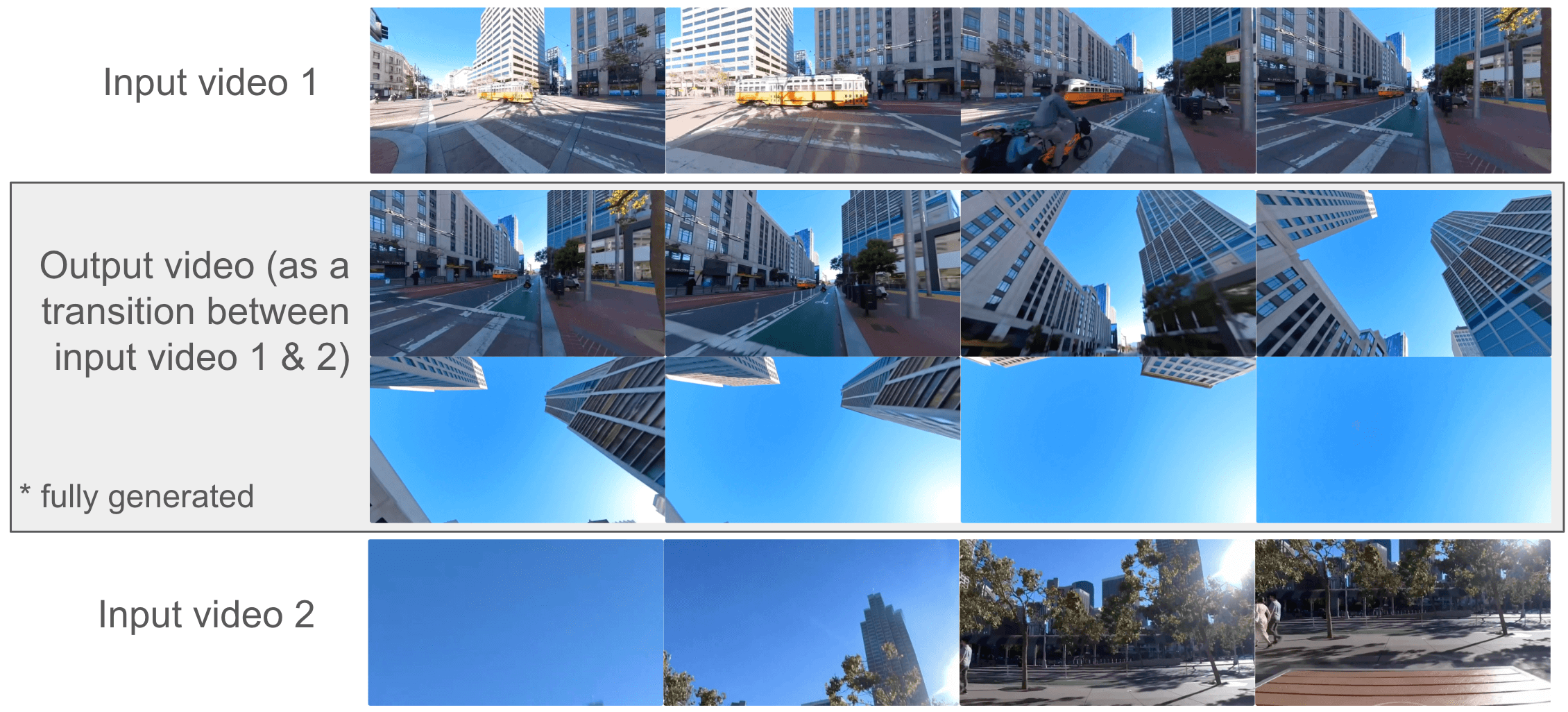}
\caption{Use case 14: Generate a smooth transition between two clips. Note that the middle two rows of transition are all 100\% generated.}
\label{fig:use_case_14}
\end{figure*}

A striking pattern was creators using text to reshoot footage long after capture by changing the viewpoint, camera motion, or scene composition. 
Participants treated Rewrite Kit as a ``virtual videographer,'' freeing them from the physical constraints of the original shoot.

For example, P1 turned a rear-facing clip of her running dog into a front-facing shot. 
She was surprised by the result:
\begin{quote}\textit{
``It's like I had a second camera in front of him---I didn't think I could manage to set it up in the real world.''}
\end{quote}
For her, text functioned as a directorial instruction: the words ``from the front, wind in its fur'' guided the system toward a shot she couldn't film alone. 
Similarly, P7 added synthetic drone views to their national park footage (as the use of drones was prohibited there). 
P3, a dancer, used text to add dynamic camera motion to a static shot, wanting to ``boost the energy of my dance.''

Expert P10, working with an animation (Figure~\ref{fig:use_case_12}), articulated how this could transform professional workflows:
\begin{quote}\textit{
``To really make an edit work, you want different angles... But it's such a hassle to get that coverage. This lets you expand any moment into a bunch of different angles you can actually cut between.''}
\end{quote}
She valued being able to describe a composition (e.g., ``low-angle shot from ground level''), edit in text prompt, and generate variations quickly, and see the result match what she ``pictured in my mind.'' 

Across these cases, text became a virtual camera-level control surface, making complex cinematography achievable through natural language specification.

\subsubsection{Text as Editorial Glue: Remixing and Bridging Footage}

Another common practice involved using text to repair, combine, or connect footage. 
Here, creators treated Rewrite Kit as an AI editor that could blend disparate clips, clean up visual obstructions, or extend scenes to improve narrative continuity.

For instance, P2 combined a wide shot of a concert with a selfie video to create an imagined moment, describing it as ``patching a memory I forgot to record.'' 
P4 used text for a restorative purpose: removing a couple who blocked the view in his speedboat footage. 
P12 extended a basketball shot to create a bridging scene for his vlog.

Expert P11 used the probe to generate seamless transitions between two otherwise disconnected clips (Figure~\ref{fig:use_case_14}), a task that typically requires meticulous planning during a shoot. 
He noted:
\begin{quote}\textit{
``Transition generation is cool. In traditional shooting, you have to plan... for a motion transition... Here, you can just generate.''}
\end{quote}
This family of practices points to what we call \textbf{synthetic continuity}: generating connective material that maintains narrative flow without manual keyframing or masking. 
With text as an interface, continuity became something creators could describe rather than manually construct.

\subsubsection{Text as a Stylist: Restyling for Vibe}

\begin{figure*}
\centering
\includegraphics[width=0.73\textwidth]{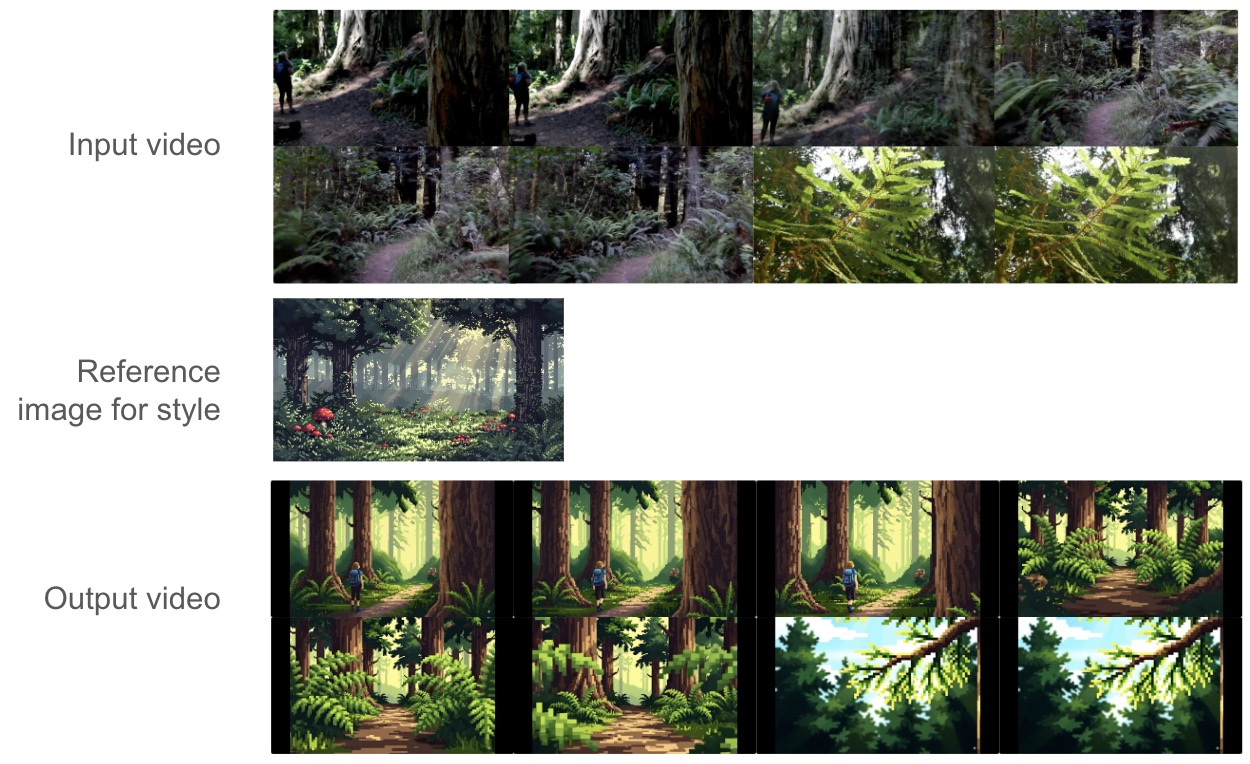}
\caption{Use case 9: Stylizing a vlog as pixel art. Guided by a single reference image, Rewrite Kit restyles the original photorealistic footage into a pixel art animation.}
\label{fig:use_case_9}
\end{figure*}

\begin{figure*}
\centering
\includegraphics[width=0.73\textwidth]{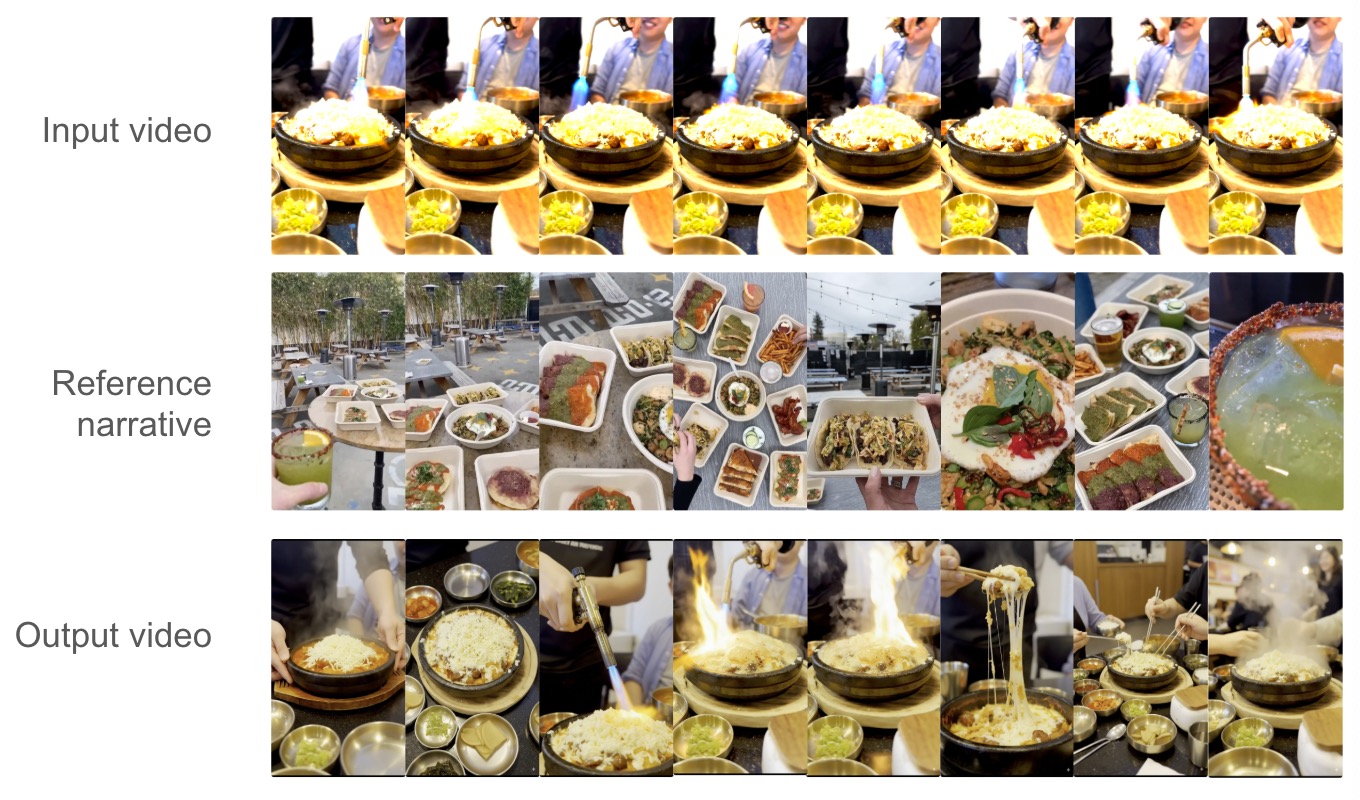}
\caption{Use case 10: Expanding a short food clip into a rich vlog. The input video, shot from a single static viewpoint, lacks narrative richness. Using a reference vlog as a guide, Rewrite Kit generates new viewpoints and synthesizes a more dynamic and engaging vlog.}
\label{fig:use_case_10}
\end{figure*}

Many participants used Rewrite Kit to transform the look, tone, or rhythm of their videos, effectively adopting the role of an art director. 
Text, often supplemented by reference videos or images, served as a vehicle for aesthetic direction.

P5, creating a YouTube intro, used a reference from a creator she admired to ``develop similar things for my own channel,'' a goal she found difficult to achieve before. 
P8 transformed her forest vlog into a pixel-art style (Figure~\ref{fig:use_case_9}), while P9 used a reference vlog to expand a simple clip of melting cheese into a richly edited, multi-angle narrative (Figure~\ref{fig:use_case_10}).

A key insight from these sessions was that the reverse-engineered prompt from a reference clip served as a ``scaffold for creativity.'' 
It provided novices with the necessary vocabulary to understand how a ``vibe'' was constructed, transforming an aesthetic they appreciated into a manipulable interface they could adapt for their own style.

\subsubsection{Text as a World-Builder: Re-writing the Scene's Reality}

\begin{figure*}
\centering
\includegraphics[width=0.745\textwidth]{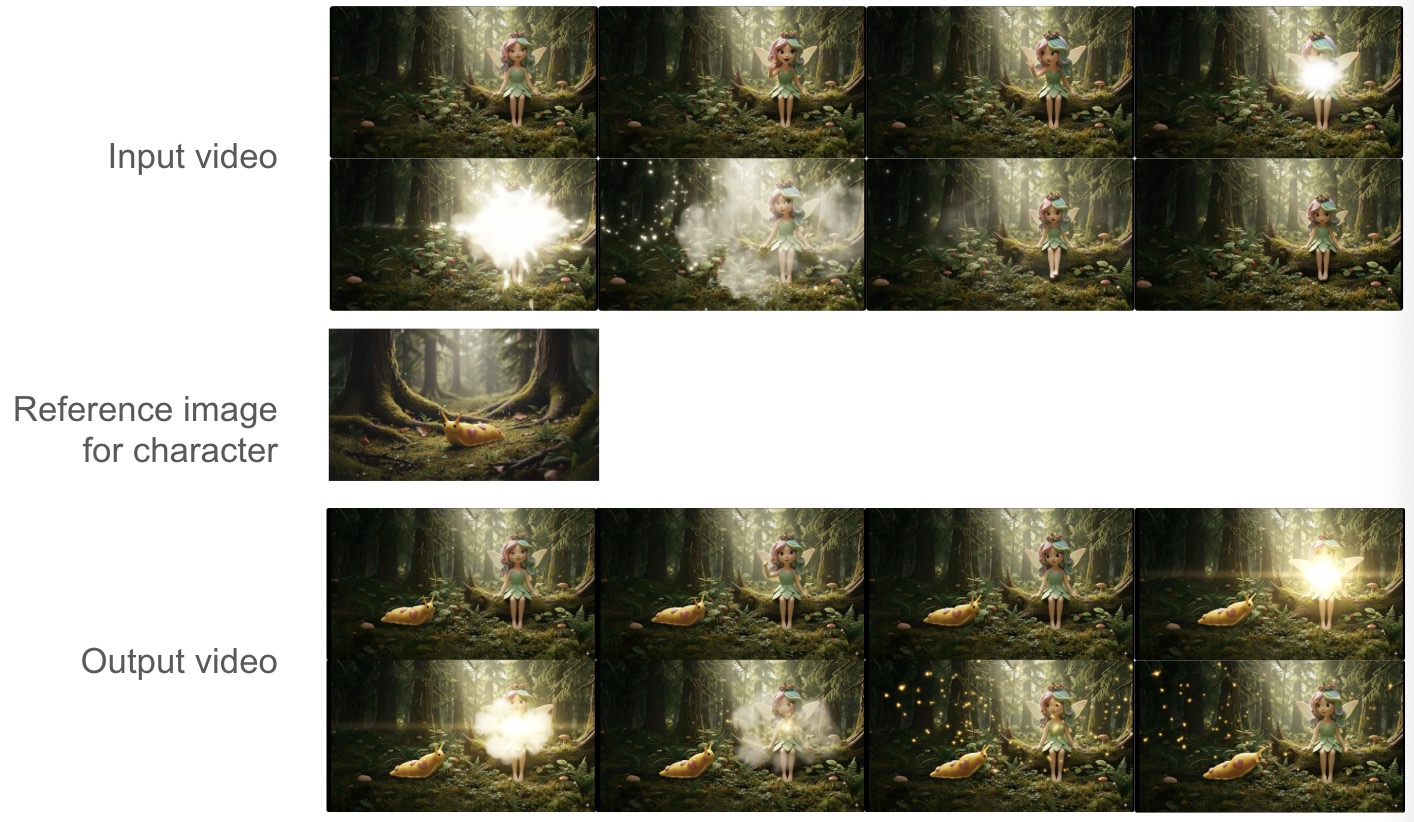}
\caption{Use case 11: Adding a yellow slug to an animation. For a participant wanting to insert a novel character from their asset library into an existing clip, Rewrite Kit provided a seamless workflow.}
\label{fig:use_case_11}
\end{figure*}

\begin{figure*}
\centering
\includegraphics[width=0.73\textwidth]{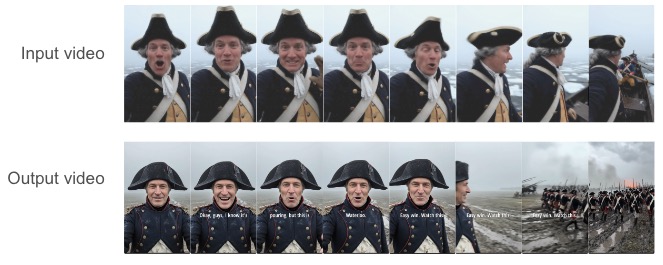}
\caption{Use case 4: Recreating a viral ``historical selfie'' with a new character. 
Inspired by a popular video of a historical figure taking a selfie, the user aimed to recreate the underlying concept with a different subject. This involved more than simple style transfer---Rewrite Kit adapted the original narrative's structure to fit this new character.
}
\label{fig:use_case_4}
\end{figure*}

Finally, a few participants approached Rewrite Kit as a world-building engine, using text to rewrite \textit{what a scene is about} rather than how it is shot. 
They inserted new characters, re-authored relationships, and reimagined worlds, transforming video editing into a mode of narrative redesign.

Expert P10 explored this by inserting a new character---a yellow slug---into her animated fairy clip (Figure~\ref{fig:use_case_11}). 
This act edited the scene's \textbf{narrative ontology}---the set of entities and events that exist within its world. 
P4 engaged in a similar practice, taking inspiration from a viral AI video of a historical ``selfie'' to restage the concept with a different figure (Figure~\ref{fig:use_case_4}). 
He reflected, ``AI makes it feel like you can rewrite history visually.''

In these cases, text enabled creators to treat videos as narrative spaces open to reinterpretation. 
The ability to modify who appears, what they do, or the world they inhabit suggests a powerful shift, transforming video from a record of captured reality into a writable, malleable universe.

\subsection{RQ2: What challenges and opportunities arise when video editing is mediated through text? What are the implications for future design?}

While text-driven editing proved powerful, it also surfaced deep tensions around coherence, control, and creative expression. 
These frictions were not simple usability errors; they were sites of tradeoffs where creators grappled with the boundaries between their intent and the model's interpretation, observed across the set of 15 use cases.
Two cases (Use Cases 3 and 5) involved partial failures due to fast-motion complexity; the remaining cases were considered successful by participants in achieving their intended goals. 
We identified four recurrent challenges that map the expressive and ethical landscape of this new paradigm.

\subsubsection{The Tension of World Coherence: Keeping the Scene ``True''}

A primary challenge was maintaining the internal logic of a scene after a generative change. 
Creators judged results not just on realism, but on whether new elements ``belonged''---whether they inherited the scene's light, physics, and atmosphere.

P10, who added a slug to her fairy animation, wanted the slug to ``sink into the moss---to feel a little more integrated.'' 
Her feedback framed coherence as a \textit{relational property} between objects, not just visual plausibility.
Similarly, P2 felt the lighting on her composited selfie was ``a bit unnatural,'' and P1 noted that without a reference photo, the AI-generated dog felt like ``a random dog,'' not her own. 
These moments highlight a new form of creative labor: \textbf{world-keeping}, or ensuring the rewritten world still ``holds together.''

\textbf{Design Opportunity:} Future systems should help creators reason about relational coherence. 
Instead of just placing objects, interfaces could allow users to define how new elements inherit world properties like illumination, physics, or causal logic, helping to sustain the scene's internal truth.

\subsubsection{The Tension of Authenticity: The Real vs. Synthetic Boundary}

Participants' tolerance for generative imperfection varied sharply depending on the source material. 
In animated or fantastical contexts, approximation was acceptable. 
But for footage of real people or cherished memories, even minor flaws felt intrusive. 
P10 articulated this asymmetry:
\begin{quote}\textit{
``I'm much more flexible when the modifications are on an animated... video... But if it were a clip of me or my kids, I'd have a higher bar for preservation.''}
\end{quote}
The most common fault line was the nuance of movement. 
P3 found the system ``not excellent at reproducing choreography,'' while others noted that a dog's jog appeared ``slightly mechanical'' or a speedboat's up-and-down motion was lost. 
These micro-failures broke the illusion of reality, reminding participants of the generative layer underneath. 
This boundary was not only aesthetic but also ethical, as creators implicitly balanced when an edit crossed from representation into fabrication.

\textbf{Design Opportunity:} Instead of chasing flawless realism, systems could introduce an “authenticity gradient”---a control that lets creators choose what kind of truth a scene should preserve: visual, emotional, or narrative. 
Making this gradient explicit would encourage reflection on where a work lies between documentary fidelity and imaginative synthesis, helping creators stay intentional about the truths their creations convey.

\subsubsection{The Tension of Translation: Aligning Creative Voice with a Reference}

Participants consistently used reference videos not to copy them, but as expressive catalysts for developing their own creative voice. 
They sought to extract and adapt specific qualities---a style, rhythm, or structure---for their own content. 
The reverse-engineered prompt often served as a ``creative schema,'' giving them the vocabulary and structure to deconstruct and reauthor an aesthetic.

However, a tension arose when the text failed to capture the essence of a reference. 
P9, working on a food vlog, felt the results were close but wanted to ``adjust the beats to match the music like the reference did,'' noting that non-visual rhythm was the heart of its appeal. 
This highlights that reauthoring is a complex act of translation and negotiation---deciding what to keep from a reference, what to transform, and how to infuse it with a personal voice. 
For P4, a viral ``George Washington selfie'' video became inspiration for restaging the idea with a different historical figure---a form of visual parody that blended homage with reinterpretation. 
He collaborated closely with the conversational AI to co-develop a script, ensuring that the new narrative remained coherent while preserving the meme's core structure. 
These examples illustrate that text-driven reauthoring is not a simple pipeline from description to output, but an ongoing negotiation of creative alignment---deciding what to keep, what to transform, and what to invent.

\textbf{Design Opportunity:} Systems could support this translation process by helping users deconstruct a reference into its core components (e.g., pacing, color palette, camera motion). 
By making these underlying principles explicit and editable, tools can help creators borrow structural ideas while developing a distinct creative voice.

\subsubsection{The Tension of Modality: When Words Are Not Enough}

Despite the power of text, participants inevitably hit expressive ceilings where language felt insufficient.
Cinematic qualities like gesture, rhythm, and composition often resisted linear, verbal description.

8 of 12 participants used the optional first-frame editing feature when it was relevant to their reauthoring goal, finding that providing a visual anchor gave them more reliable control. 
Others expressed a desire for more visual and direct manipulation. 
P8 wanted a storyboard to preview keyframes before the ``heavy video generation process,'' while P10 wished she could ``directly annotate on the frame and tell the model what happens next.'' 
When collaborating with the AI, many participants naturally began constructing multimodal prompts, combining images and text to convey their intent more precisely.

\textbf{Design Opportunity:} The ``perfect prompt'' is a myth. 
A good prompt is often a combination of text, image, and cues in other modalities. 
Future systems should treat text as one channel within a broader multimodal conversation. 
Interfaces that allow creators to fluidly interleave describing, showing, and marking will better support the embodied, iterative nature of creative thought.
\section{Discussion}

Our study illustrates how text-driven video reauthoring reconfigures creative work, shifting it from frame-level manipulation toward high-level semantic creation with generative models. 
While the preceding sections identified empirical patterns, here we discuss the broader implications of our findings for creative agency, authorship, and the design of future co-creation tools.

\subsection{From Direct Manipulation to Semantic Generation}
Text-driven interfaces transform video reauthoring from a process of manual control into semantic writing and generation. 
Rather than manipulating pixels on a timeline, creators articulate their intent through language and iteratively steer the system toward a desired outcome. This reframes authorship as a conversational act centered on expressing, interpreting, and refining intent with an AI partner. 
This model does not replace the HCI tradition of direct manipulation~\cite{shneiderman1983direct}, but complements it with a paradigm grounded in semantic expression and shared agency. 
Future co-creation tools should therefore prioritize interpretability—making system reasoning visible and editable—over simply providing precise control of low-level parameters.
We observed that prior editing experience appeared to shape engagement: expert participants tended to favor lower-level technical edits (e.g., camera changes or transitions), while novices were more open to higher-level narrative manipulation (e.g., expanding a short clip into a full vlog).

\subsection{The Perceptual Gap as a Design Material}

Our evaluation revealed a persistent \textbf{human-AI perceptual gap}: while models optimize for frame-level fidelity, human judgment hinges on temporal rhythm, emotional tone, and narrative coherence. 
Rather than viewing this gap as a technical flaw to be eliminated, it can be treated as a design material for reflection. 
Future interfaces could expose the system’s interpretation of ``similarity,'' allowing creators to consciously decide when to prioritize visual accuracy versus narrative flow, depending on their own taste and preference. 
Such transparency would elevate generative tools from opaque assistants to reflective collaborators, supporting more intentional control over the temporal and affective meaning of a video.

\subsection{World-Keeping: A New Form of Creative Labor}
Generative transformations introduced a distinct form of creative work we term \textbf{world-keeping}: the labor of sustaining the internal logic, physics, and atmosphere of the source material while rewriting its content. 
A successful edit was not defined by realism alone but by whether it ``belonged''—whether its light, motion, and mood cohered with the original scene. 
While this world-keeping will improve as generative models become more capable, we believe it is critical to develop interfaces that make relational coherence an explicit, editable property, enabling creators to specify what dimensions of reality (visual, emotional, causal) they wish to preserve, instead of offloading all the decisions to the opaque models. 
In this sense, world-keeping marks a new locus of creative authorship: the craft of maintaining continuity in an inherently fluid, generative medium.

\subsection{Learning Through Re-creation with Reference Media}
Creators consistently used reference clips not to copy, but to learn how a particular ``vibe'' is constructed. 
By reverse-engineering prompts from examples, they gained a scaffold for understanding the constituent elements of a style—pacing, camera work, color—which they then re-contextualized in their own work. 
This process treats exemplars as analyzable, decomposable models of style rather than as fixed templates. 
Future systems could make this deconstruction explicit, breaking references into editable layers (e.g., rhythm, tone, motion dynamics) to better support creative learning and reinterpretation over one-click style transfer.

\subsection{Beyond Text: Toward Multimodal Authoring}

Although text afforded powerful high-level control, participants repeatedly turned to visual and temporal cues—first frames, examples, or imagined storyboards—when language reached its expressive limits. 
This highlights the need for multimodal authoring environments where text, image, gesture, and timing operate as complementary channels of intent. 
As underlying models become increasingly multimodal, the design challenge shifts to orchestrating these inputs fluidly, allowing creators to describe with words, demonstrate with visuals, and mark through temporal cues within a single iterative loop.

\section{Limitations and Future Work}

First, our probe study focused on short video clips (typically under 10 seconds), reflecting the current capabilities of text-to-video models. 
Future work should explore how these text-driven practices scale to longer narratives, which may require new techniques for maintaining global continuity across decomposed scenes.

Second, while Rewrite Kit is positioned to explore high-level narrative and structural reauthoring, most use cases in our probe study involved adapting existing narratives with support from reference media. 
Such scaffolding may make reauthoring more accessible---particularly for novice creators---but limits our ability to assess how effectively users can invent substantially new narrative content through text alone. 
Exploring more open-ended, reference-free narrative reauthoring remains an important direction for future work.

Third, this work is situated within a rapidly evolving technological landscape. 
We did not evaluate our system with the most recent models (e.g., Sora 2~\cite{sora2025openai}) since it came out only days before the paper submission deadline. 
While we expect our core design insights around control, intent alignment, and coherence to hold, future studies should revisit these ideas with next-generation architectures.

Finally, our prototype prioritized a text-first workflow.
Integrating other emerging modalities, such as annotation-based or sketch-based video manipulation, could yield richer, more embodied forms of control while preserving the conceptual clarity of text-driven interaction.

\paragraph{Ethical Considerations}
Text-driven reauthoring amplifies critical ethical concerns. 
The ease of rewriting footage blurs boundaries of authorship and can be used to copy a creator's stylistic identity~\cite{porquet2025copying} or manipulate real-world imagery in deceptive ways~\cite{nyt_ai_videos_2025}. 
Addressing these risks is both a technical challenge and a design imperative. 
Future systems should be built on a foundation of responsibility, embedding safeguards such as provenance tracking~\cite{C2PA_Spec_2_2}, style attribution, and consent indicators to foster a culture of transparent and ethical creative practice.

\section{Conclusion}
In this paper, we investigated a new paradigm for video creation: reauthoring footage as fluidly as rewriting text.
Through our generative reconstruction algorithm and the \textit{Rewrite Kit} probe, we demonstrate that exposing a video's underlying textual ``script'' empowers creators to reshape narratives through language, enabling practices like virtual reshooting, restyling, and remixing that move beyond timeline constraints. 
Our findings reveal both the promise and the friction of this approach: while text enables powerful, intent-driven control, it also surfaces fundamental challenges around coherence, multimodal expression, and creative alignment.
Ultimately, this work reframes video reauthoring as an act of expressive writing, pointing toward a future of co-creative systems that make visual storytelling more fluid, interpretable, and fundamentally more accessible.

%\section{Generative AI Usage Disclosure}
%Generative AI tools were used solely to refine the authors’ own writing for clarity and grammar. 
%No generative AI tools were used for project ideation, study design, data analysis, or result interpretation.

\bibliographystyle{ACM-Reference-Format}
\bibliography{sample-base}

@inproceedings{guo2025keyframe,
  title={Keyframe-Guided Creative Video Inpainting},
  author={Guo, Yuwei and Yang, Ceyuan and Rao, Anyi and Meng, Chenlin and Bar-Tal, Omer and Ding, Shuangrui and Agrawala, Maneesh and Lin, Dahua and Dai, Bo},
  booktitle={Proceedings of the Computer Vision and Pattern Recognition Conference},
  pages={13009--13020},
  year={2025}
}

@misc{runway_creatingwithaleph,
  title        = {Creating with Aleph},
  author       = {{Runway ML}},
  howpublished = {\url{https://help.runwayml.com/hc/en-us/articles/43176400374419-Creating-with-Aleph}},
  year         = {n.d.},
  note         = {Accessed: 2025-10-07},
  organization = {Runway ML Help Center}
}

@article{xie2025reconstruction,
  title={Reconstruction alignment improves unified multimodal models},
  author={Xie, Ji and Darrell, Trevor and Zettlemoyer, Luke and Wang, XuDong},
  journal={arXiv preprint arXiv:2509.07295},
  year={2025}
}

@article{wiedemer2025video,
  title={Video models are zero-shot learners and reasoners},
  author={Wiedemer, Thadd{\"a}us and Li, Yuxuan and Vicol, Paul and Gu, Shixiang Shane and Matarese, Nick and Swersky, Kevin and Kim, Been and Jaini, Priyank and Geirhos, Robert},
  journal={arXiv preprint arXiv:2509.20328},
  year={2025}
}

@inproceedings{zhu2017unpaired,
  title={Unpaired image-to-image translation using cycle-consistent adversarial networks},
  author={Zhu, Jun-Yan and Park, Taesung and Isola, Phillip and Efros, Alexei A},
  booktitle={Proceedings of the IEEE international conference on computer vision},
  pages={2223--2232},
  year={2017}
}

@inproceedings{hutchinson2003technology,
  title={Technology probes: inspiring design for and with families},
  author={Hutchinson, Hilary and Mackay, Wendy and Westerlund, Bo and Bederson, Benjamin B and Druin, Allison and Plaisant, Catherine and Beaudouin-Lafon, Michel and Conversy, St{\'e}phane and Evans, Helen and Hansen, Heiko and others},
  booktitle={Proceedings of the SIGCHI conference on Human factors in computing systems},
  pages={17--24},
  year={2003}
}

@inproceedings{huber2019b,
  title={B-script: Transcript-based b-roll video editing with recommendations},
  author={Huber, Bernd and Shin, Hijung Valentina and Russell, Bryan and Wang, Oliver and Mysore, Gautham J},
  booktitle={Proceedings of the 2019 CHI Conference on Human Factors in Computing Systems},
  pages={1--11},
  year={2019}
}

@inproceedings{rubin2013content,
  title={Content-based tools for editing audio stories},
  author={Rubin, Steve and Berthouzoz, Floraine and Mysore, Gautham J and Li, Wilmot and Agrawala, Maneesh},
  booktitle={Proceedings of the 26th annual ACM symposium on User interface software and technology},
  pages={113--122},
  year={2013}
}

@article{fried2019text,
  title={Text-based editing of talking-head video},
  author={Fried, Ohad and Tewari, Ayush and Zollh{\"o}fer, Michael and Finkelstein, Adam and Shechtman, Eli and Goldman, Dan B and Genova, Kyle and Jin, Zeyu and Theobalt, Christian and Agrawala, Maneesh},
  journal={ACM Transactions on Graphics (TOG)},
  volume={38},
  number={4},
  pages={1--14},
  year={2019},
  publisher={ACM New York, NY, USA}
}

@article{roughcut,
  title={Computational video editing for dialogue-driven scenes.},
  author={Leake, Mackenzie and Davis, Abe and Truong, Anh and Agrawala, Maneesh},
  journal={ACM Trans. Graph.},
  volume={36},
  number={4},
  pages={130--1},
  year={2017}
}

@inproceedings{podreels,
  title={PodReels: Human-AI Co-Creation of Video Podcast Teasers},
  author={Wang, Sitong and Ning, Zheng and Truong, Anh and Dontcheva, Mira and Li, Dingzeyu and Chilton, Lydia B},
  booktitle={Proceedings of the 2024 ACM Designing Interactive Systems Conference},
  pages={958--974},
  year={2024}
}

@inproceedings{radford2021learning,
  title={Learning transferable visual models from natural language supervision},
  author={Radford, Alec and Kim, Jong Wook and Hallacy, Chris and Ramesh, Aditya and Goh, Gabriel and Agarwal, Sandhini and Sastry, Girish and Askell, Amanda and Mishkin, Pamela and Clark, Jack and others},
  booktitle={International conference on machine learning},
  pages={8748--8763},
  year={2021},
  organization={PmLR}
}

@misc{sora2025openai,
  title        = {Sora: Creating Video from Text},
  author       = {OpenAI},
  year         = {2025},
  howpublished = {\url{https://openai.com/index/sora/}},
  note         = {Accessed: 2025-10-07}
}

@misc{runway2025gen4,
  title        = {Introducing Runway Gen-4},
  author       = {Runway Research},
  year         = {2025},
  howpublished = {\url{https://runwayml.com/research/introducing-runway-gen-4}},
  note         = {Accessed: 2025-10-07}
}

@inproceedings{brade2023promptify,
  title={Promptify: Text-to-image generation through interactive prompt exploration with large language models},
  author={Brade, Stephen and Wang, Bryan and Sousa, Mauricio and Oore, Sageev and Grossman, Tovi},
  booktitle={Proceedings of the 36th Annual ACM Symposium on User Interface Software and Technology},
  pages={1--14},
  year={2023}
}

@inproceedings{lee2022coauthor,
  title={Coauthor: Designing a human-ai collaborative writing dataset for exploring language model capabilities},
  author={Lee, Mina and Liang, Percy and Yang, Qian},
  booktitle={Proceedings of the 2022 CHI conference on human factors in computing systems},
  pages={1--19},
  year={2022}
}

@inproceedings{wang2024promptcharm,
  title={Promptcharm: Text-to-image generation through multi-modal prompting and refinement},
  author={Wang, Zhijie and Huang, Yuheng and Song, Da and Ma, Lei and Zhang, Tianyi},
  booktitle={Proceedings of the 2024 CHI Conference on Human Factors in Computing Systems},
  pages={1--21},
  year={2024}
}

@inproceedings{liu2022design,
  title={Design guidelines for prompt engineering text-to-image generative models},
  author={Liu, Vivian and Chilton, Lydia B},
  booktitle={Proceedings of the 2022 CHI conference on human factors in computing systems},
  pages={1--23},
  year={2022}
}

@article{dang2022prompt,
  title={How to prompt? Opportunities and challenges of zero-and few-shot learning for human-AI interaction in creative applications of generative models},
  author={Dang, Hai and Mecke, Lukas and Lehmann, Florian and Goller, Sven and Buschek, Daniel},
  journal={arXiv preprint arXiv:2209.01390},
  year={2022}
}

@article{braun2006using,
  title={Using thematic analysis in psychology},
  author={Braun, Virginia and Clarke, Victoria},
  journal={Qualitative research in psychology},
  volume={3},
  number={2},
  pages={77--101},
  year={2006},
  publisher={Taylor \& Francis}
}

@inproceedings{porquet2025copying,
  title={Copying style, Extracting value: Illustrators' Perception of AI Style Transfer and its Impact on Creative Labor},
  author={Porquet, Julien and Wang, Sitong and Chilton, Lydia B},
  booktitle={Proceedings of the 2025 CHI Conference on Human Factors in Computing Systems},
  pages={1--16},
  year={2025}
}

@misc{google2025gemini,
  author       = {{Google AI}},
  title        = {Generate videos with Veo 3 in Gemini API: prompt guide},
  howpublished = {\url{https://ai.google.dev/gemini-api/docs/video?example=dialogue\#prompt-guide}},
  note         = {Last updated 2025-10-06 (UTC)},
  year         = {2025}
}

@article{shneiderman1983direct,
  title={Direct manipulation: A step beyond programming languages},
  author={Shneiderman, Ben},
  journal={Computer},
  volume={16},
  number={08},
  pages={57--69},
  year={1983},
  publisher={IEEE Computer Society}
}

@manual{C2PA_Spec_2_2,
  title        = {C2PA Specification, Version 2.2},
  organization = {Coalition for Content Provenance and Authenticity (C2PA)},
  year         = {2025},
  url          = {https://spec.c2pa.org/specifications/specifications/2.2/index.html},
  note         = {Accessed: 7 October 2025}
}

@inproceedings{boehner2007hci,
  title={How HCI interprets the probes},
  author={Boehner, Kirsten and Vertesi, Janet and Sengers, Phoebe and Dourish, Paul},
  booktitle={Proceedings of the SIGCHI conference on Human factors in computing systems},
  pages={1077--1086},
  year={2007}
}

@article{nyt_ai_videos_2025,
  title        = {A.I. Videos Have Never Been Better. Can You Tell What’s Real?},
  author       = {{The New York Times}},
  journal      = {The New York Times},
  year         = {2025},
  month        = {June 29},
  url          = {https://www.nytimes.com/2025/06/29/technology/ai-videos-real-or-fake.html},
  note         = {Accessed: 2025-10-07}
}

\end{document}